\documentclass[twocolumn,showpacs,preprintnumbers,amsmath,amssymb,aps]{revtex4-2}
\usepackage{graphicx}
\usepackage{dcolumn}
\usepackage{bm}
\usepackage[colorlinks=true]{hyperref}
\usepackage{epstopdf}
\usepackage[T1]{fontenc} % special letters
\usepackage{lmodern}
\usepackage{braket}
\usepackage{ulem}
\usepackage{color}

\definecolor{color}{rgb}{0.11,0.45,0.02}

\renewcommand{\d}{\mathrm d}

\begin{document}

\title{Magnetic field induced exciton spin dynamics in indirect band gap (In,Al)As/AlAs quantum dots}

\author{T.~S.~Shamirzaev$^{1}$, D.~R.~Yakovlev$^{2,3}$, D.~S.~Smirnov$^{3}$, V.~N.~Mantsevich$^{4}$, D.~Kudlacik$^{2}$, A.~Yu.~Gornov$^{5}$, A.~K.~Gutakovskii$^{1}$, and M.~Bayer$^{2,6}$}

\affiliation{
$^1$Rzhanov Institute of Semiconductor Physics, Siberian Branch of the Russian Academy of Sciences, 630090 Novosibirsk, Russia \\
$^2$Experimentelle Physik 2, Technische Universit\"at Dortmund, 44227 Dortmund, Germany \\
$^3$Ioffe Institute, Russian Academy of Sciences, 194021 St.
Petersburg, Russia \\
$^4$Lomonosov Moscow State University, 119991, Moscow, Russia \\
$^5$Institute for System Dynamics and Control Theory, Siberian Branch of the Russian Academy of Sciences, 664033 Irkutsk, Russia \\
$^6$ Research Center FEMS, Technische Universit\"at Dortmund, 44227 Dortmund, Germany}

\begin{abstract}
The exciton recombination and spin dynamics are investigated both experimentally and theoretically in an ensemble of indirect band gap (In,Al)As/AlAs quantum dots (QDs) with type-I band alignment. The magnetic-field-induced circular polarization of the time-integrated photoluminescence changes sign across the emission spectrum with a width reflecting the QD size. It is negative on the low energy side, i.e. for emission from large QDs, but positive on the high energy side, i.e. for emission from small QDs. However, the exciton $g$ factor, measured by spin-flip Raman scattering, is positive across the whole QD ensemble. The magnetic-field-induced circular polarization of the photoluminescence dynamics is studied as function of the magnetic field strength and direction. The dynamics are non-monotonic over a time range up to milliseconds. The time dependence of the photoluminescence circular polarization degree and sign strongly depends on the emission energy and changes with magnetic field orientation. The observed nonmonotonic behavior is provided by the interplay of bright and dark exciton states, contributing to the emission. The experiment is interpreted using a kinetic theory, which accounts for the dynamics of the spin states in the exciton level quartet in longitudinal and tilted magnetic fields, the radiative recombination processes, and the redistribution of the excitons between these states as result of spin relaxation. The model allows us to evaluate the electron and heavy hole spin relaxation times in QDs with different sizes.
\end{abstract}

\maketitle
\section{Introduction}
\label{sec:intro}

Semiconductor quantum dots (QDs) have been applied in industrial fields ranging from electronics to optical products such as solar cells, photovoltaic devices, and single-photon sources~\cite{Klingshirn,Somaschi}. One of the interesting features of QDs are the long spin lifetimes of electrons, holes, and excitons~\cite{Khaetskii0,Kroutvar,Glazov} that are of great relevance for spin-based quantum information processing~\cite{Burkard,Dyakonov}. A fundamental parameter of excitons confined in QDs is their recombination time, which can be adjust by the type of band gap~\cite{Shamirzaev94}. The exciton lifetime in direct band gap QDs is rather short, on the order of a nanosecond~\cite{Yu}, while the recombination dynamics of excitons in QDs with indirect band gap are extremely long lasting up to several hundreds of microseconds at cryogenic temperatures~\cite{Abramkin112},  which  allows  one  to  study  long  living spin dynamics~\cite{Ivchenko60,Shamirzaev60}. It was demonstrated that indirect in momentum space band gap QDs with  type-I  band alignment can be realised in an (In,Al)As/AlAs heterostructure~\cite{Shamirzaev78,Jacobson}. The exciton spin dynamics in small magnetic fields are  determined mainly by the electron-nuclear hyperfine interaction~\cite{Smirnov125,Shamirzaev13o,Kuznetsova,Shamirzaev104nn} and electron-hole exchange  interaction~\cite{Rautert100,Nekrasov}, as was shown for large size (In,Al)As/AlAs QDs. This allows for obtaining non-equilibrium spin polarization via optical orientation under quasi-resonant excitation~\cite{Rautert99}.

Unfortunately, optical orientation does not work in small size indirect band gap QDs with large energy gap between the ground X and excited $\Gamma$ electron states due to spin relaxation during energy relaxation. However, there is a convenient technique to address the dynamics of electron and exciton spins localized in
indirect band gap QDs, namely studying the redistribution of populations between the spin states split by a magnetic field, which manifests itself in the magnetic field-induced circular polarization~\cite{Ivchenko60}. The redistribution of exciton populations
between the Zeeman states leads to circular polarization of the photoluminescence (PL), induced by the magnetic field.

Nontrivial dependencies of the circular polarization degree ($P_c$) on time and magnetic field were theoretically predicted for excitons with long lifetimes~\cite{Shamirzaev96}. In this paper, we aim at demonstrating them experimentally. We investigate the effect of a magnetic field on the exciton recombination and spin dynamics in an ensemble of indirect band gap (In,Al)As/AlAs QDs with type-I band alignment. Indeed, we find that inspite of the identical sign of the bright exciton $g$ factor for the whole indirect band gap QD ensemble, $P_c$ shows a nomonotonic spectral dependence with a change of polarization sign across the PL emission band. $P_c$ also demonstrates a nonmonotonic dynamics with a change in polarization sign, which  depends on  the  magnetic filed strength and orientation. These experimental signatures can be explained by a kinetic model of the exciton dynamics.

The paper is organized as follows. In Sec.~\ref{sec:details} the studied samples and  experimental techniques are described. In
Sec.~\ref{sec:results} we present the experimental data on the exciton dynamics obtained in an external magnetic field by time-integrated and
time-resolved PL. Also the time-integrated and time-resolved circular polarization of the PL induced by the magnetic field are given. The theoretical model describing the QD exciton dynamics is presented in Sec.~\ref{sec:discussion}. The experimental data are analyzed in the frame of this model. The conclusions are given in Sec.~\ref{sec:conclusions}.

\section{Experimental details}
\label{sec:details}

The studied self-assembled (In,Al)As QDs, embedded in an AlAs matrix, were grown by molecular-beam epitaxy on semi-insulating (001)-oriented GaAs substrates.
The structure contains three QD sheets sandwiched between 50-nm-thick AlAs layers grown on top of a 400-nm-thick GaAs buffer layer. The nominal amount of deposited InAs was about 2.55 monolayers per sheet. A 20-nm-thick GaAs cap layer protected the AlAs layer against oxidation.
Further details for the epitaxial growth of QDs in the AlAs matrix are given in Ref.~\cite{Shamirzaev78}. In order to increase the exciton lifetime the structure was annealed for 10 min at the elevated temperature $T_{an} = 735^\circ$C~\cite{Shamirzaev84}. To prevent surface decomposition during the annealing, the sample was protected by a 150~nm thick SiO$_2$ layer. Technical details can be found in Refs.~\cite{Shamirzaev3,Shamirzaev13n}.

\begin{figure}[]
\includegraphics* [width=8 cm]{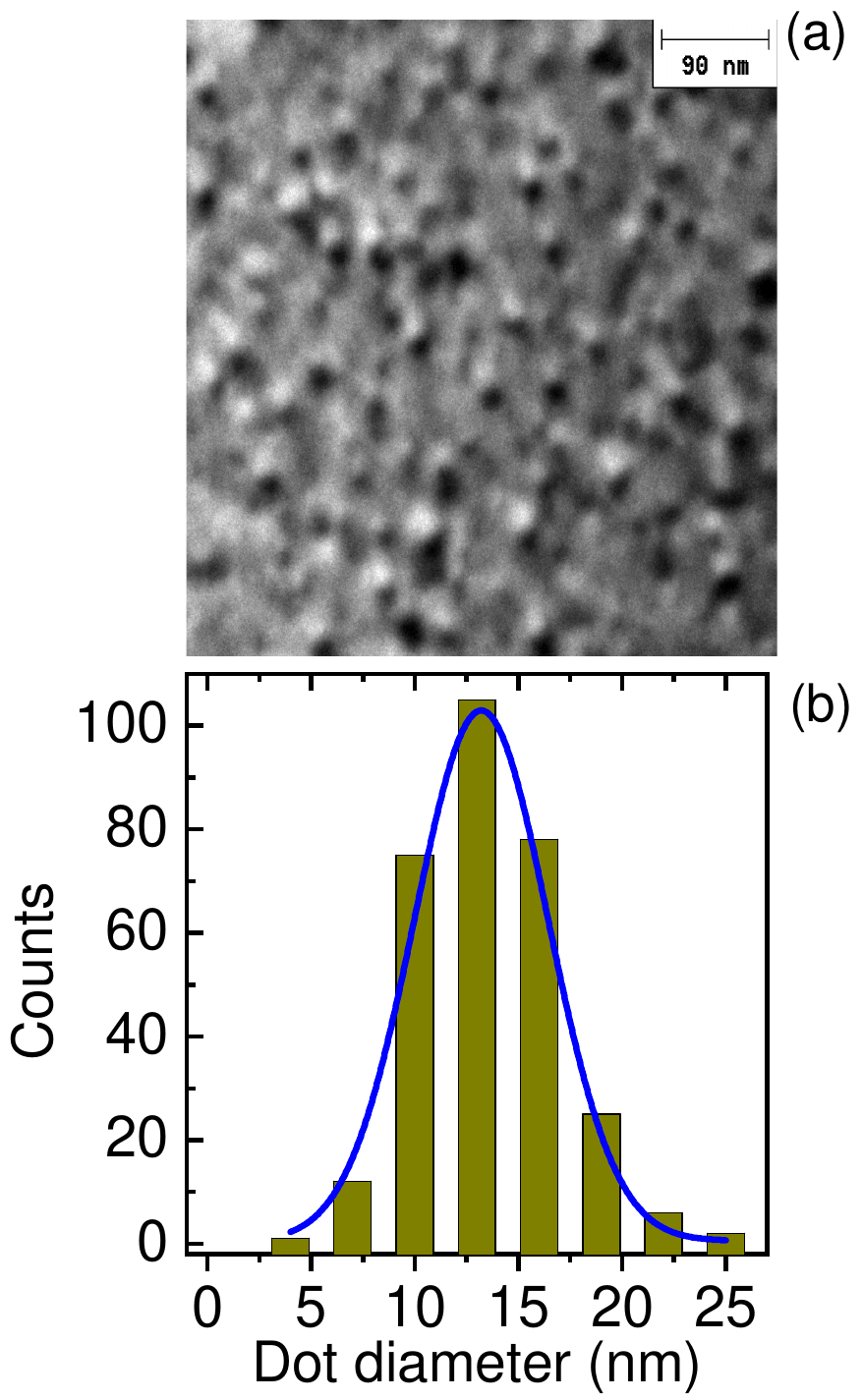}
\caption{\label{fig1} (a) TEM plane-view image of (In,Al)As/AlAs QD sample and (b) histogram of the diameter distribution of (In,Al)As/AlAs QDs.}
\end{figure}

The QD size and density were studied by transmission electron microscopy (TEM) using a JEM-4000EX system operated at an acceleration voltage of 250 keV. From the TEM images we find that the self-assembled (In,Al)As QDs are lens shaped with a typical aspect ratio (height to diameter) of about 1:4~\cite{ShamirzaevAPL92}.
A TEM plane-view image and the respective histogram of the QD-diameter distribution are shown in Fig.~\ref{fig1}. The density of the QDs in each layer is about $5.2\times10^{10}$~cm$^{-2}$. The average diameters $D_{av}$ (13.1 nm) and the diameters corresponding to the larger $D_L$ (15.8 nm) and smaller $D_S$ (8.2 nm) half-widths of the QD-size distribution are obtained from the TEM. Additionally, the size dispersion $S_D$, which is defined by the ratio of the half-width of the Gaussian distribution of QD sizes to the average diameter $S_D = 100 \% \times(D_L -D_S)/D_{av} = 57 \% $ is calculated. Since the shape of the PL emission reflects the distribution of QD sizes the average QD composition can be evaluated from  comparison of the PL band energy position with the results of model calculations~\cite{Shamirzaev78}. The determined QD composition is In$_{0.6}$Al$_{0.4}$As. Note that the relatively low QD density prevents carrier redistribution between the QDs~\cite{ShamirzaevNT,ShamirzaevSST}.

The sample is mounted strain free on a rotatable stage, which is immersed in pumped liquid helium at the temperature of $T=1.8$~K and subjected to magnetic fields up to $B=10$~T in a split-coil magnet cryostat. The angle $\theta$  between the magnetic field $\mathbf{B}$ and the QD growth $z$-axis varies from 0$^\circ$ (Faraday geometry) up to 45$^\circ$.

For the resonant spin-flip Raman scattering (SFRS) measurements, the scattered light is analyzed by a 1-m focal-length double monochromator equipped with a cooled GaAs photomultiplier providing a spectral resolution of about 10~$\mu$eV. For excitation, a tunable continuous-wave Ti:Sapphire laser is used with a power density at the sample of  15 W/cm$^{2}$. The SFRS spectra are measured in the backscattering geometry with circular or linear polarization for the incident and scattered light~\cite{Debus}.

The exciton spin dynamics were analyzed from the PL by measuring the circular polarization degree $P_c$ induced by the external magnetic field. $P_c$ was evaluated from the data by:
\begin{align}
P_c = \frac{I_{\sigma^+} - I_{\sigma^-}}{I_{\sigma^+} +
I_{\sigma^-}},\nonumber
\end{align}
where $I_{\sigma^+}$ and $I_{\sigma^-}$ are the intensities of the $\sigma^{+}$ and $\sigma^{-}$ polarized PL components, respectively.
To determine the sign of $P_c$, we performed a control measurement on a diluted magnetic semiconductor structure with
(Zn,Mn)Se/(Zn,Be)Se quantum wells for which $P_c>0$ (i.e., the stronger PL component corresponds to $\sigma^{+}$-polarized emission) in the Faraday geometry~\cite{Keller}.

The time-integrated and time-resolved PL measurements were performed at the temperature of $T = 1.8$~K. The PL was exited by the third harmonic of a Q-switched Nd:YVO$_4$ laser (3.49~eV photon energy) with a pulse duration of 5~ns. The pulse-repetition frequency was varied from 1 to~100 kHz and the pulse energy density was kept below 100 nJ/cm$^{2}$, which corresponds to about 30$\%$ probability of QD occupation with a single exciton~\cite{Shamirzaev84}. For the time-resolved PL spectra measurements we used a gated charge-coupled device camera, synchronized with the laser via an external trigger signal. The time between the pump pulse and the start of PL recording, $t_{del}$, could be varied from 0 up to 100~$\mu$s. The duration of PL recording, i.e., the gate window $t_{gate}$, could be varied from 5~ns to 100~$\mu$s. The signal intensity and the time resolution of the setup depend on $t_{del}$ and $t_{gate}$. The PL dynamics at a selected energy was detected by a GaAs photomultiplier operating in the time-correlated photon-counting mode. In order to monitor the PL decay in a wide temporal range up to 1~ms, the time resolution of the detection system was varied between 3.2~ns and 1.5~$\mu$s.

\section{Experimental results}
\label{sec:results}

\subsection{Unpolarized photoluminescence}

\subsubsection{Steady state PL in zero magnetic field}

\begin{figure}[t]
\includegraphics* [width=7.0cm]{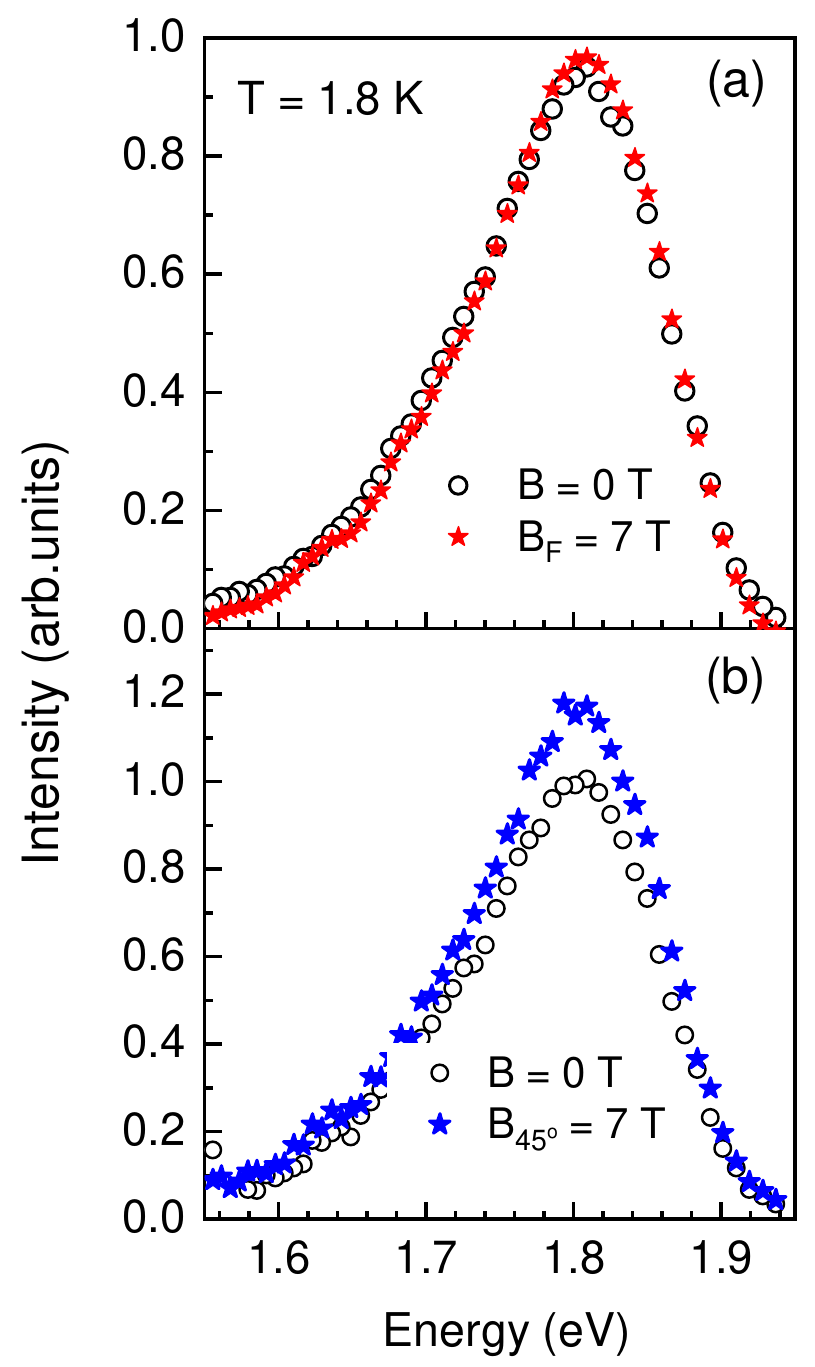}
\caption{\label{fig2} Time-integrated, unpolarized PL spectra of (In,Al)As/AlAs QDs in zero magentic field and magnetic field of 7~T applied in: (a) Faraday geometry ($\theta$ = 0$^\circ$) and (b) tilted geometry ($\theta$ = 45$^\circ$).}
\end{figure}

A time-integrated, unpolarized photoluminescence spectrum of the (In,Al)As/AlAs QDs,
measured at the temperature of 1.8~K with a laser pulse repetition rate
of 100~kHz and excitation density of 100~mW/cm$^{2}$ in zero magnetic field, is shown in Fig.~\ref{fig2}(a). The time-integrated spectrum (black open circles) has a maximum at 1.80~eV and extends from 1.55 to 1.95~eV having a full width at half maximum (FWHM) of 210~meV. The large width of the emission band is due to the dispersion of the QD parameters, since the exciton energy depends on the QD size, shape, and composition~\cite{Shamirzaev78,Shamirzaev84}.

\subsubsection{Dynamics of PL in zero magnetic field}

\begin{figure}[]
\includegraphics* [width=7.9cm]{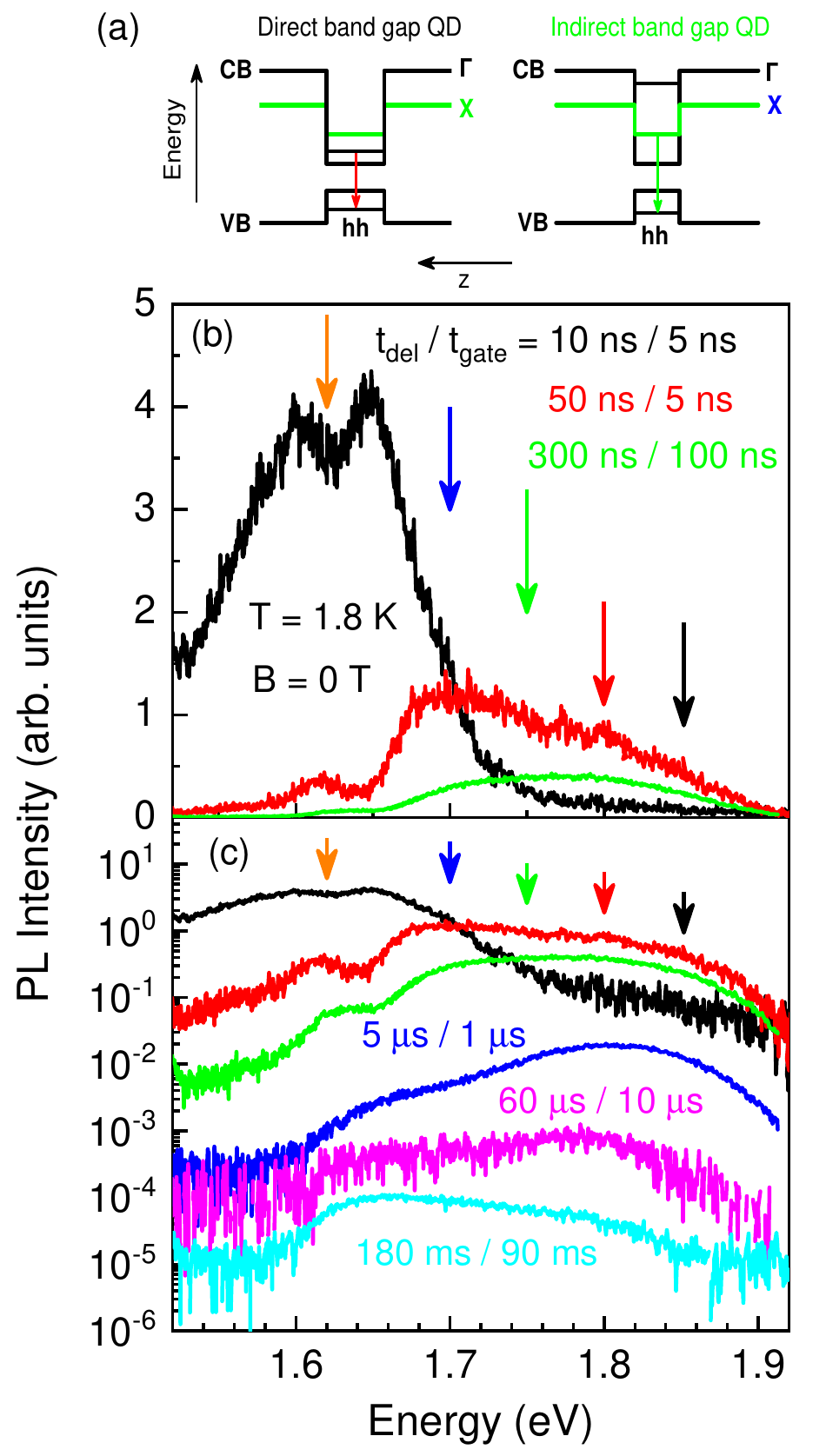}
\caption{\label{fig3}(a) Schematic diagrams of type-I (In,Al)As/AlAs QDs with direct and indirect band structure. Arrows mark the optical transitions related to the radiative decay of the ground state exciton. (b,c) Time-resolved PL spectra of (In,Al)As/AlAs QDs measured at $T = 1.8$ K for non-resonant excitation (pulsed laser with photon energy of 3.49~eV). The end of the excitation pulse corresponds to 10~ns time. (b) Linear scale and (c) Logarithmic scale for t$_{del}$/t$_{gate}$: black (10~ns/5~ns), red (50~ns/5~ns), green (300~ns/100~ns), blue (5~$\mu$s/1~$\mu$s), magenta (60~$\mu$s/10~$\mu$s), and cyan (180~$\mu$s/90~$\mu$s).  Arrows mark the energies of 1.85, 1.80, 1.75, 1.70, and 1.61~eV where the PL dynamics shown in Fig.~\ref{fig4} were measured.}
\end{figure}

Time-resolved, unpolarized PL spectra measured for different delays after the excitation pulse $t_{del}$ are shown in Figs.~\ref{fig3}(b) and~\ref{fig3}(c). One can see that the spectrum shape changes strongly with time. As they were measured immediately after the laser pulse application, at delays of several nanoseconds the PL is dominated by a low energy band $1.55-1.70$~eV related to recombination in direct band gap QDs~\cite{Rautert99}. For delay times of tens of microseconds, the high-energy range $1.75-1.85$~eV dominates in the PL spectra. And, finally, for delay times of several hundred microseconds, the middle range of the PL spectrum at $1.65-1.75$~eV dominates. Schematic diagrams of type-I QDs with direct and indirect band structure are shown in Fig.~\ref{fig3}(a). The electron ground state changes from the $\Gamma$ to the X valley with decreasing dot size, while the heavy-hole ground state remains at the $\Gamma$ point~\cite{Shamirzaev78,Rautert99}.

\begin{figure}[]
\includegraphics* [width=7.5cm]{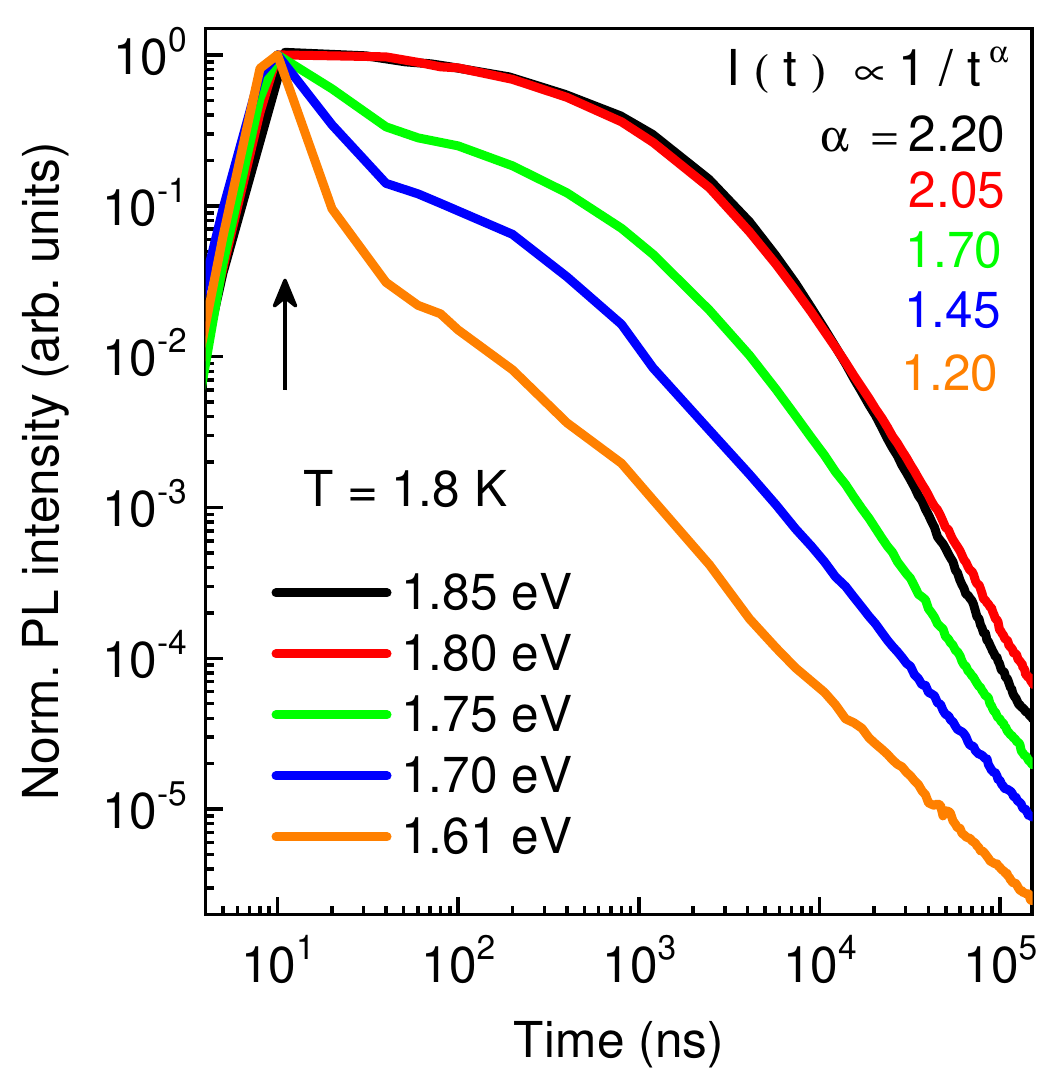}
\caption{\label{fig4} PL dynamics measured at detection energies of 1.85~eV (black), 1.80~eV (red), 1.75~eV (green), 1.70~eV (blue), and 1.61~eV (orange).}
\end{figure}

The dynamics of exciton recombination in zero magnetic field along the PL band (at energies that are indicated by arrows in Figs.~\ref{fig3}(b) and~\ref{fig3}(c)) are shown in Fig.~\ref{fig4}. The transient PL data are plotted using a double-logarithmic scale, which is convenient to illustrate the nonexponential character of the decay over a wide range of times and PL intensities. In the high-energy range of the spectrum $\geq 1.80$~eV the recombination dynamics demonstrates two distinctive stages: (i) a relatively flat PL decay immediately after the excitation pulse and (ii) a reduction of the PL intensity, which can be described by a power-law function $I(t)\propto 1/t^{\alpha}$, as shown in our previous studies~\cite{Shamirzaev84}. The non-exponential character of the PL decay  is due to the dispersion of the lifetimes of excitons emitting at the same wavelength, but localized in QDs of different sizes and compositions~\cite{Shamirzaev84}. In the low-energy part of the spectrum $1.55-1.75$~eV, in the initial stage of the decay curve an additional fast decay (10-30 nanoseconds duration) appears. This decay is due to the contribution of direct-band-gap QDs, that coexist with indirect-band-gap QDs in the ensemble. Their relative contribution to the PL increases as the proportion of the direct-band-gap fration in the QD ensemble increases.

\subsubsection{Effect of magnetic field}

\begin{figure}[]
\includegraphics* [width=8.6cm]{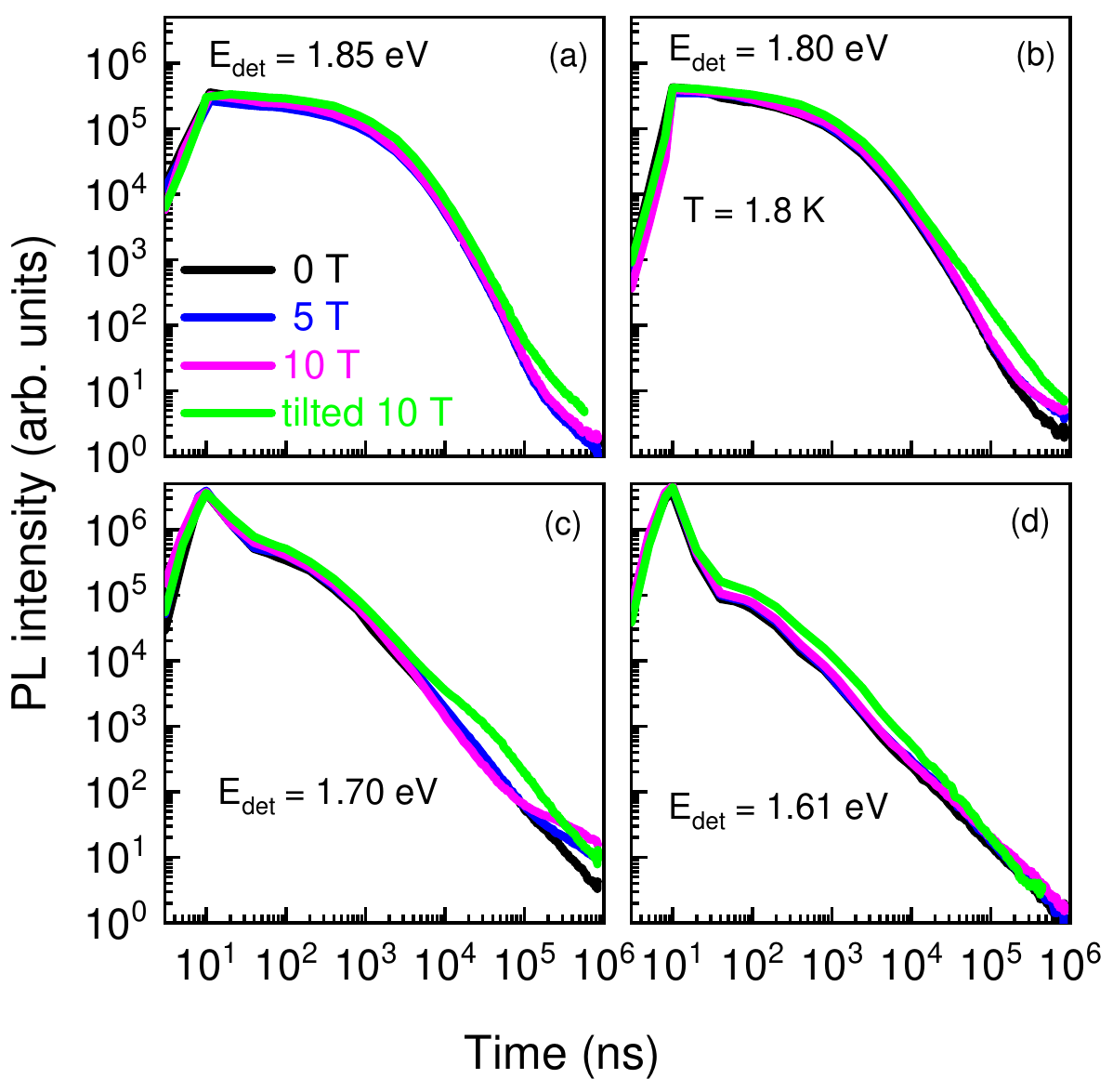}
\caption{\label{fig5} Unpolarized PL dynamics measured in magnetic fields (Faraday geometry) of 0, 5, and 10~T, and at 10~T in tilted geometry ($\theta=35^\circ$, green curves) at the detection energies of  (a) 1.85~eV, (b) 1.80~eV, (c) 1.70~eV, and (d) 1.61~eV. The excitation pulse ends at 10~ns.}
\end{figure}

A longitudinal magnetic field does not affect the unpolarized PL intensity in the time-integrated spectra, while a tilted magnetic field ($\theta=45^\circ$) leads to an increase of the PL intensity by about 20${\%}$, as it is shown in Figs.~\ref{fig2}(a) and~\ref{fig2}(b).

The spectral dependence of the unpolarized PL dynamics in different magnetic fields are shown in Figs.~\ref{fig5}. In longitudinal magnetic fields up to 10~T, the PL dynamics in all spectral points are basically unchanged up to a delay time of 100~$\mu$s after the excitation pulse. For longer times, a slowdown of the PL decay is observed which becomes more pronounced with increasing field in the energy range of $1.70 - 1.80$~eV.  On the other hand, for energies above 1.80~eV and below 1.70~eV the PL dynamics do not depend on the magnetic field strength in the entire range of measurement times.

When the magnetic field is tilted to $\theta$ = 35$^\circ$, there is some slowdown of the PL decay for delays exceeding 10~$\mu$s in the high-energy range above 1.65~eV.

\subsection{Magnetic field-induced circular polarization}

\subsubsection{Steady state PL}
\begin{figure}[]
\includegraphics* [width=7.2cm]{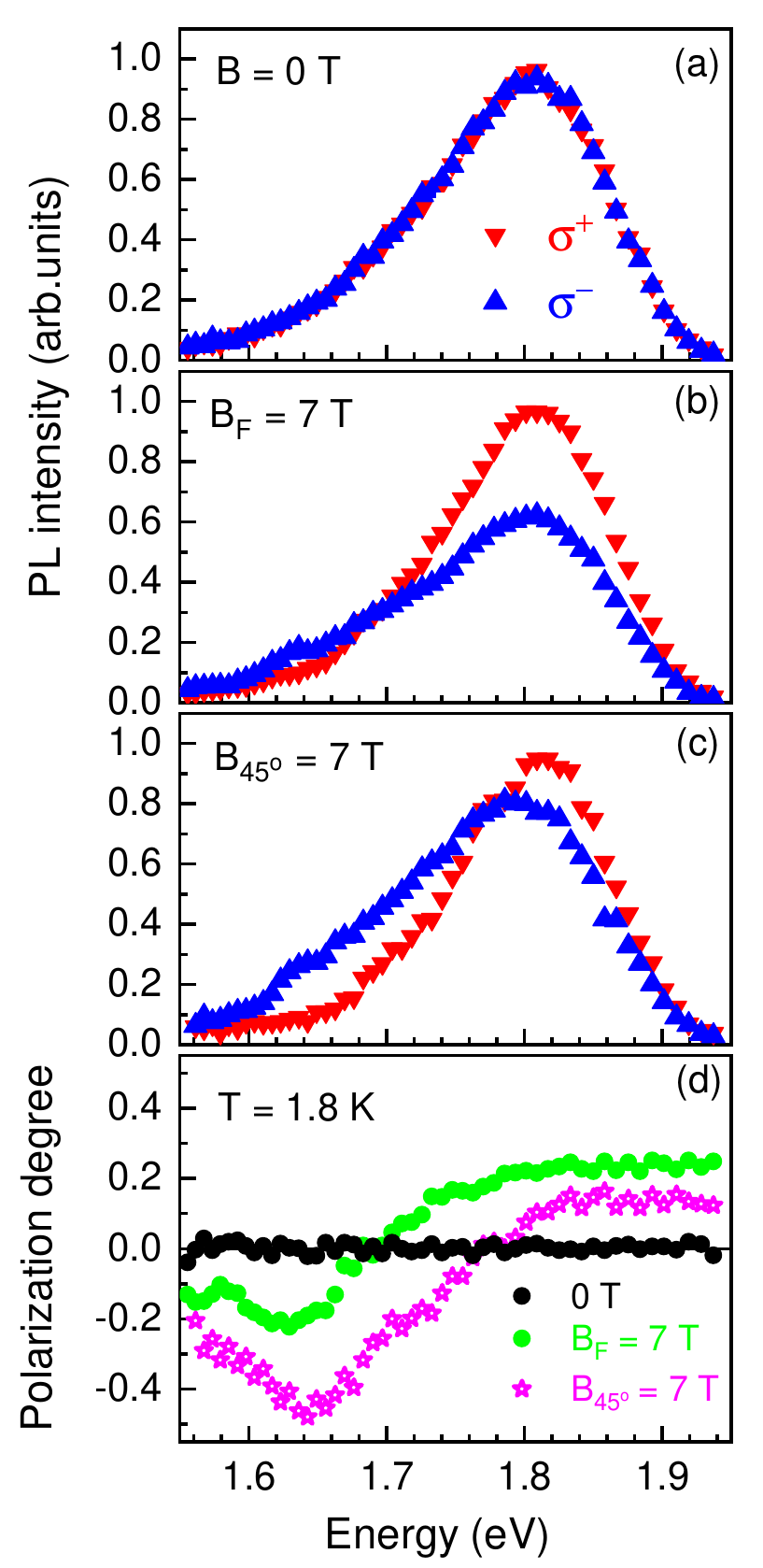}
\caption{\label{fig6} Time-integrated PL spectra measured for opposite circular polarizations $\sigma^{+}$ (red) and $\sigma^{-}$ (blue) in magnetic fields of: (a) 0~T, (b) 7~T Faraday geometry, (c) 7~T tilted to $\theta$ = 45$^\circ$, (d) spectral dependencies of the PL circular polarization degree in zero (black), 7~T longitudinal (green), and 7~T tilted to $\theta$ = 45$^\circ$ (magenta) magnetic field. $T=1.8$~K.}
\end{figure}

The time-integrated PL spectra of the (In,Al)As/AlAs QDs measured for opposite circular polarizations $\sigma^{+}$ and $\sigma^{-}$ at zero field as well as in magnetic field of 7~T applied in Faraday geometry ($\theta=0^\circ$) and tilted geometry  ($\theta=45^\circ$) are shown in Figs.~\ref{fig6}(a),~\ref{fig6}(b), and~\ref{fig6}(c), respectively. In zero magnetic field, the PL is unpolarized: the $\sigma^{+}$ and $\sigma^{-}$ polarized PL components have the same intensity. Application of a magnetic field in the Faraday geometry results in polarization of the emission as shown in Figs.~\ref{fig6}(b) and~\ref{fig6}(c).
The polarization is not uniform across the spectrum. In longitudinal fields, it is positive (i.e., it is dominated by the $\sigma^{+}$  polarized PL component) in the spectral range  $1.67-1.95$~eV, negative in the spectral range $1.56-1.67$~eV, and tends to zero for energies smaller than $1.56$~eV. In tilted geometry, the spectral dependence of the PL circular polarization changes. The spectral range corresponding to positive polarization narrows to $1.77-1.95$~eV, while the one with negative polarization expands to $1.56-1.77$~eV.

\subsubsection{Time-resolved PL}

\begin{figure}[]
\includegraphics* [width=9cm]{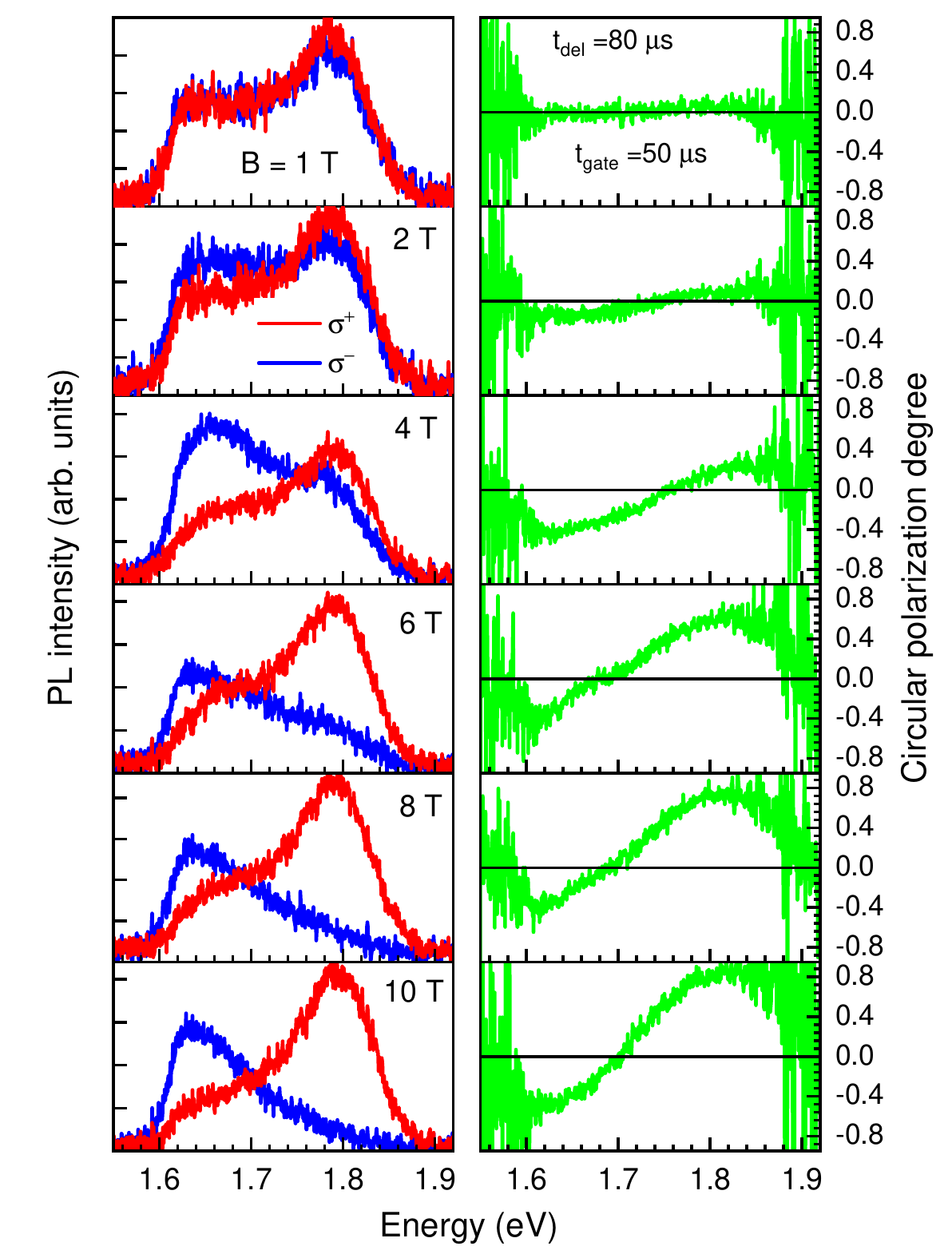}
\caption{\label{fig7} (Left panels) Time-resolved PL spectra of (In,Al)As/AlAs QDs measured in  $\sigma^{+}$ (red) and $\sigma^{-}$ (blue) circular polarization for different magnetic fields. (Right panels) Corresponding spectral dependencies of the MCPL degree. Faraday geometry, $t_{del}$/$t_{gate}$ = 80~$\mu$s/50~$\mu$s, $T=1.8$~K. From top to bottom the magnetic field strength is 1, 2, 4, 6, 8, and 10~T.}
\end{figure}

The spectral non-uniformity of the magnetic-field-induced polarization becomes more obvious in time resolved measurements. Time-resolved PL spectra measured for $\sigma^{+}$ and $\sigma^{-}$ circular polarization at the delay $t_{del} = 80~\mu$s in the time window $t_{gate}=50~\mu$s are shown in Fig.~\ref{fig7} for different magnetic fields. One can see that the high magnetic field spectra can be clearly divided in two counter-polarized parts.

\begin{figure}[]
\includegraphics* [width=8.4cm]{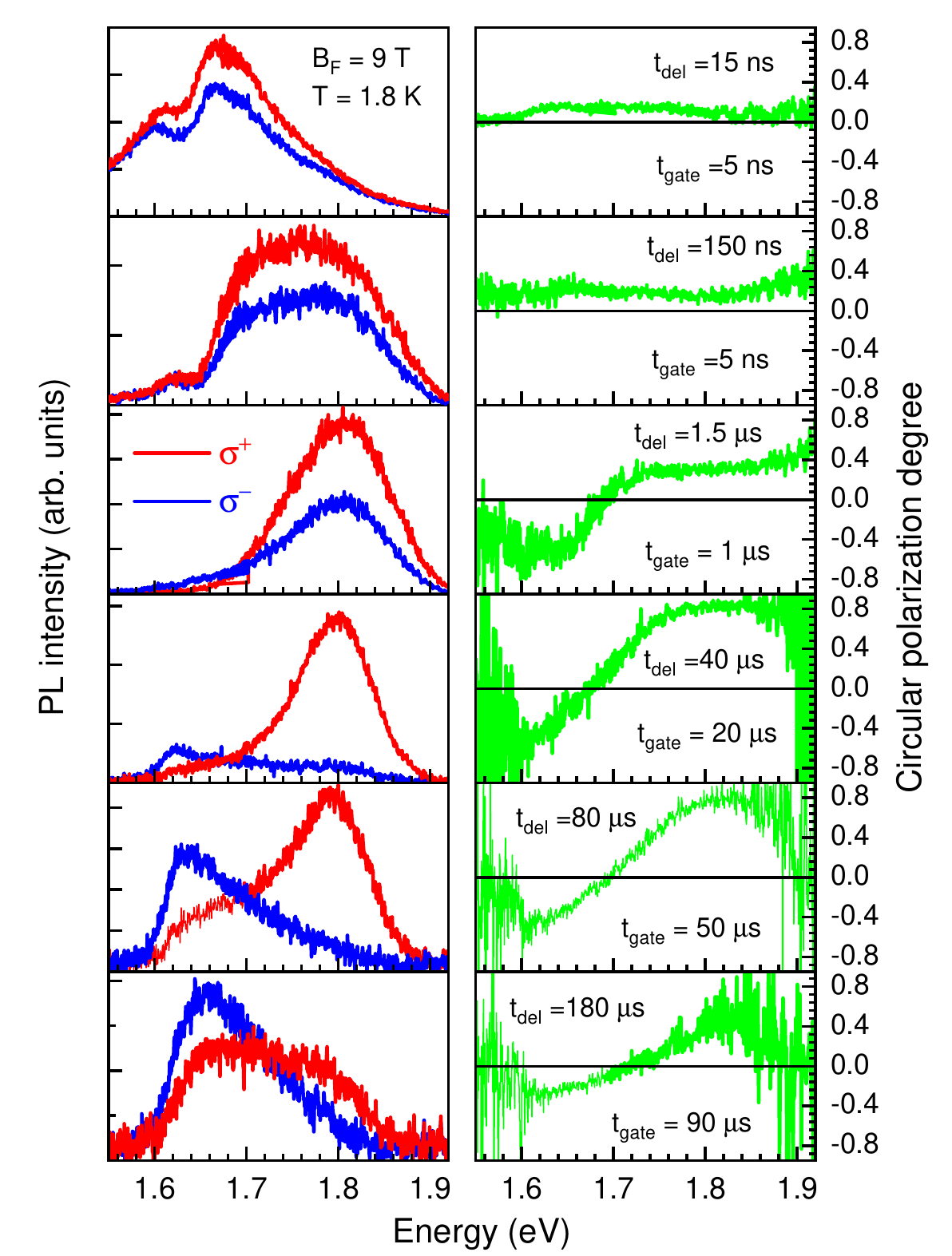}
\caption{\label{fig8} (Left panels) Time-resolved PL spectra of (In,Al)As/AlAs QDs, measured in opposite circular polarizations $\sigma^{+}$ (red) and $\sigma^{-}$ (blue). (Right panels) Corresponding spectral dependencies of the MCPL degree. $B_F=9$~T in Faraday geometry at $T=1.8$~K. From top to bottom, the ratio $t_{del}$/$t_{gate}$ changes as: (15~ns/5~ns), (150~ns/5~ns), (1500~ns/100~ns), (40~$\mu$s/20~$\mu$s), (80~$\mu$s/50~$\mu$s), and (180~$\mu$s/90~$\mu$s).}
\end{figure}

In order to more closely look into the $P_{c}$ spectral dependence, we measured the time-resolved PL at different delays after the excitation pulse. Time-resolved PL spectra measured in $\sigma^{+}$ and $\sigma^{-}$ circular polarization in a longitudinal magnetic field of 9~T (Faraday geometry) for different delays after the excitation pulse, $t_{del}$, are shown in Fig.~\ref{fig8} in the left panels. The corresponding spectral dependencies of the magnetic-field-induced PL circular polarization (MCPL) degree are shown in the right panels. Here, one should highlight a feature that is independent of the exciton recombination energy (QD size): immediately after the excitation pulse ($t_{del} \approx 10$~ns), the  polarization is positive with a degree of about +0.2 and changes only slightly up to delays $t_{del}$ of several hundred nanoseconds. At $t_{del}$ of about a microsecond, the polarization remains positive  in the high-energy range of the spectrum ($> 1.7$~eV), but becomes negative in the low-energy range ($<1.7$~eV). The polarization degrees, both positive and negative, increase with increasing delay, $t_{del}$, up to several tens of microseconds, after which they begin to decrease.

\begin{figure}[]
\includegraphics* [width=8.9cm]{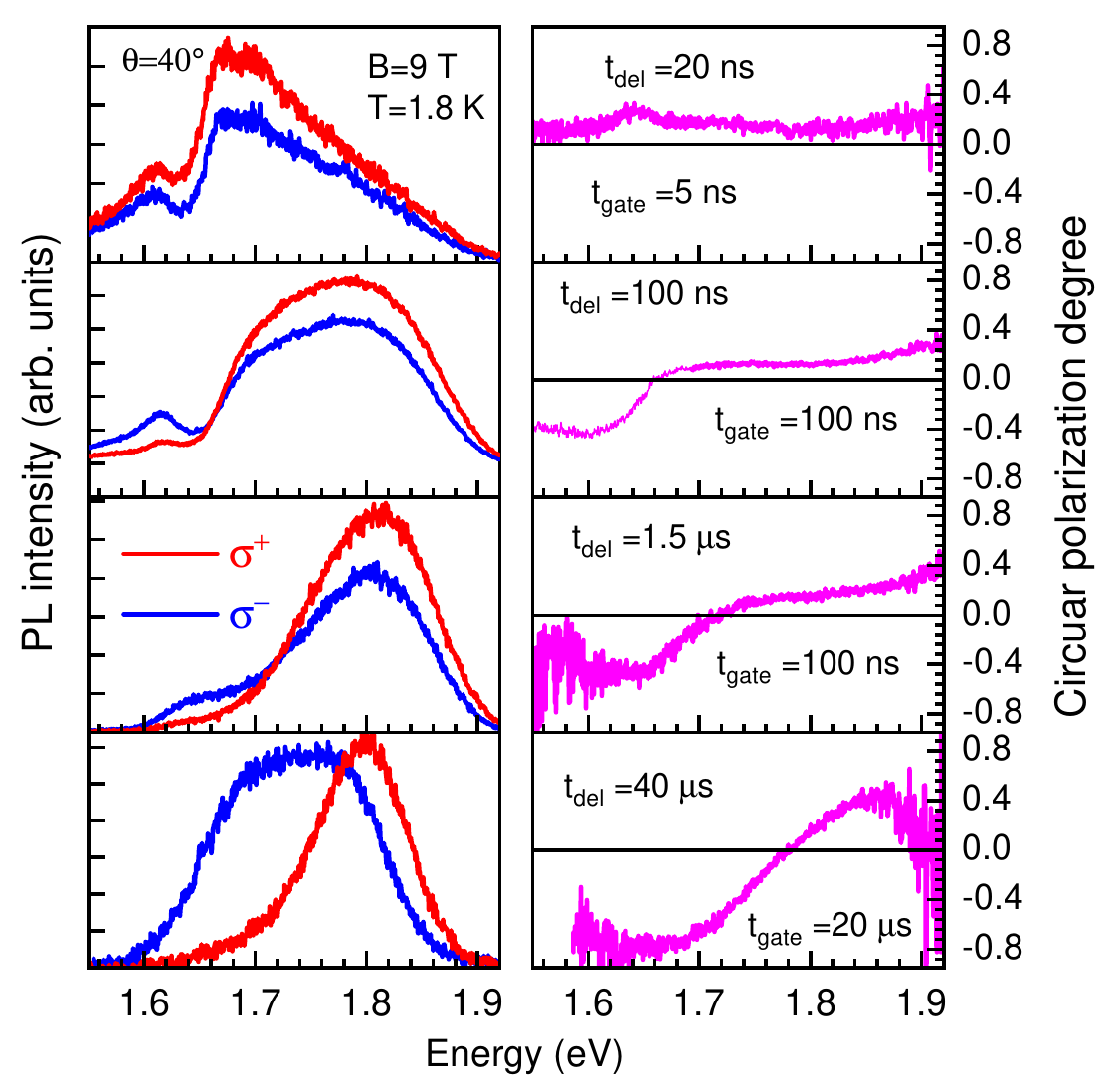}
\caption{\label{fig9} (Left panels) Time-resolved PL spectra of (In,Al)As/AlAs QDs measured in $\sigma^{+}$ (red) and $\sigma^{-}$ (blue) circular polarization. (Right panels)  Corresponding spectral dependencies of the MCPL degree. Tilted geometry ($\theta$ = 40$^\circ$), $B=9$~T at $T=1.8$~K. From top to bottom, the ratio $t_{del}$/$t_{gate}$ changes as: (20~ns/5~ns), (100~ns/100~ns), (1500~ns/100~ns), and (40~$\mu$s/20~$\mu$s).}
\end{figure}

Tilting of the magnetic field ($\theta$ = 40$^\circ$) does not change the spectral dependence of the PL polarization (the polarization is positive and does not depend on the exciton recombination energy) at delays $t_{del}$ up to about 100~ns, see Fig.~\ref{fig9}. However, the negative polarization at the low-energy edge of the spectrum appears already for $t_{del} = 100$~ns (i.e. in tilted field, the spin relaxation of excitons emitting in this spectral range accelerates).

\subsubsection{Dynamics of circular polarized PL}

\begin{figure}[]
\includegraphics* [width=8.5cm]{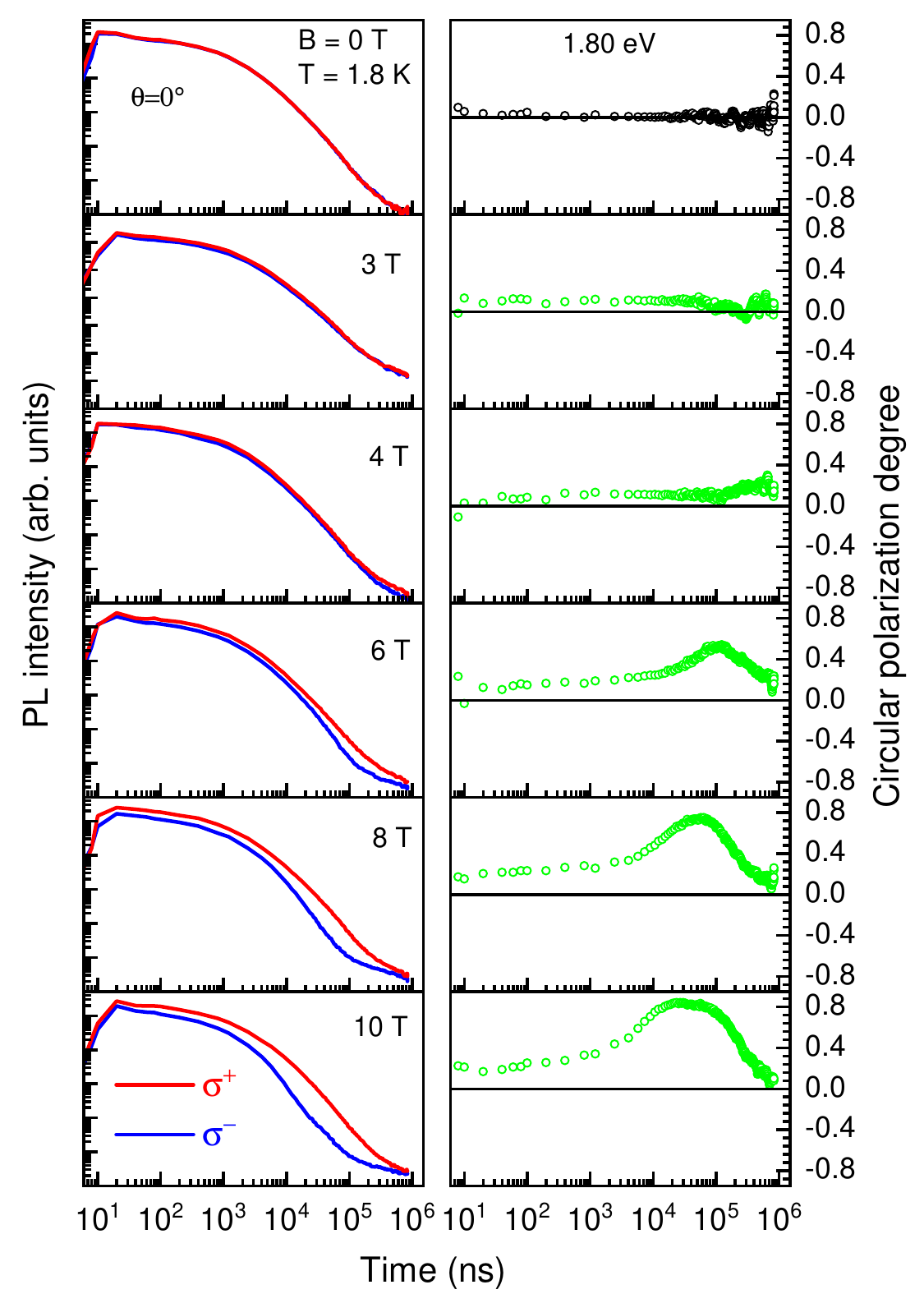}
\caption{\label{fig10} (Left panels) PL dynamics measured in $\sigma^{+}$ (red) and $\sigma^{-}$ (blue) circular polarization in different magnetic fields. (Right panels) Corresponding dynamics of the MCPL degree. Faraday geometry ($\theta= 0 ^\circ$) at $T=1.8$~K, PL detection energy of 1.80~eV.}
\end{figure}

In order to take a closer look into the spin dynamics, we measured the PL dynamics in $\sigma^{+}$ and $\sigma^{-}$ circular polarization across the PL band (i.e., for QD ensembles exhibiting positive and negative circular polarization in the PL spectra) following non-resonant pulsed excitation with linearly polarized light (to avoid optical orientation of charge carriers). The measurements were taken in magnetic fields of different strength (0 up to 10~T) and orientations ($\theta=$ 0$^\circ$ and 35$^\circ$).

With increasing longitudinal field ($\theta=$ 0$^\circ$) strength from 0 up to 4~T, the PL circular polarization shows up right after the end of the excitation pulse and does not depend on time up to 1~ms, see Fig.~\ref{fig10}. A further increase of the field strength (from 6 up to 10~T), the dependence $P_{c}(t)$ becomes non-monotonic. It increases over tens of microseconds to a certain maximum positive value  $P^{max}_{c}$, that depends on the field strength ($P^{max}_{c} = + 0.84$ at $B = 10$~T) and begins to decrease almost to zero at times exceeding a hundred microseconds.
\begin{figure}[]
\includegraphics* [width=8.9cm]{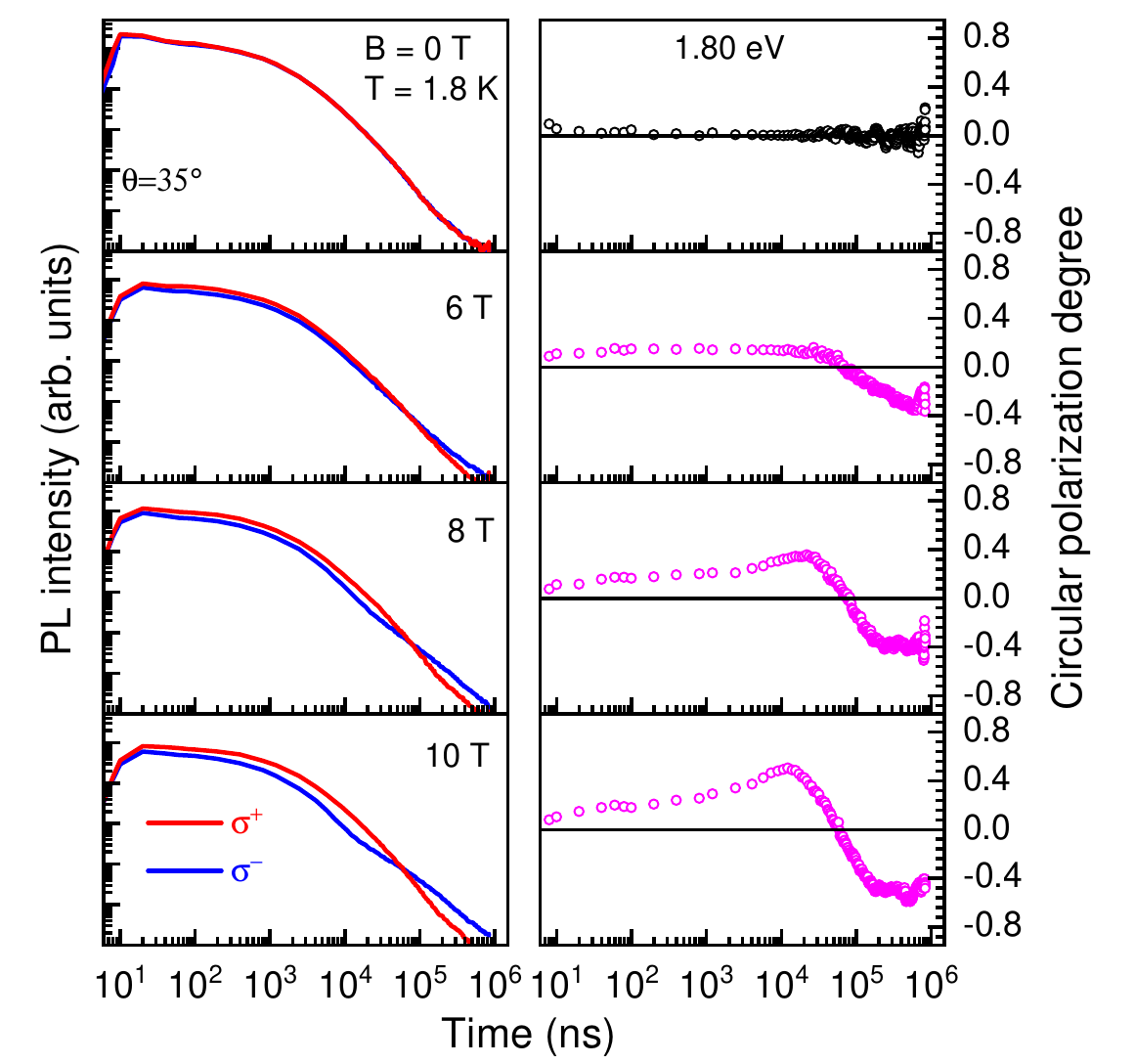}
\caption{\label{fig11}(Left panels) PL dynamics measured in $\sigma^{+}$ (red) and $\sigma^{-}$ circular polarization (blue) in  various magnetic fields. (Right panels) Corresponding dynamics of the MCPL degree. Tilted geometry ($\theta=$ 35$^\circ$), $T=1.8$~K, PL detection energy of 1.80~eV.}
\end{figure}
Tilting the magnetic field by $\theta=$ 35$^\circ$ does not effect the $P_c$ dynamics up to fields of 4~T ($P_c(t)$ coincides with that observed in Faraday geometry). However, in stronger tilted fields the picture changes qualitatively, as shown in Fig.~\ref{fig11}. In a field of 6~T, the polarization firstly is positive and constant, but within several tens of microseconds it begins to decrease to zero and then becomes negative increasing in absolute value. For a further increase in magnetic field strength, $P_c(t)$ is positive and rises during the first ten microseconds up to a certain maximum value ($P_c = +0.5$ at 12~$\mu$s in 10~T) and then over time decreases to zero and changes sign, continuously increasing in absolute value until saturation ($P_c = -0.5$ at 1~ms in 10~T).

\begin{figure}[]
\includegraphics* [width=8.8cm]{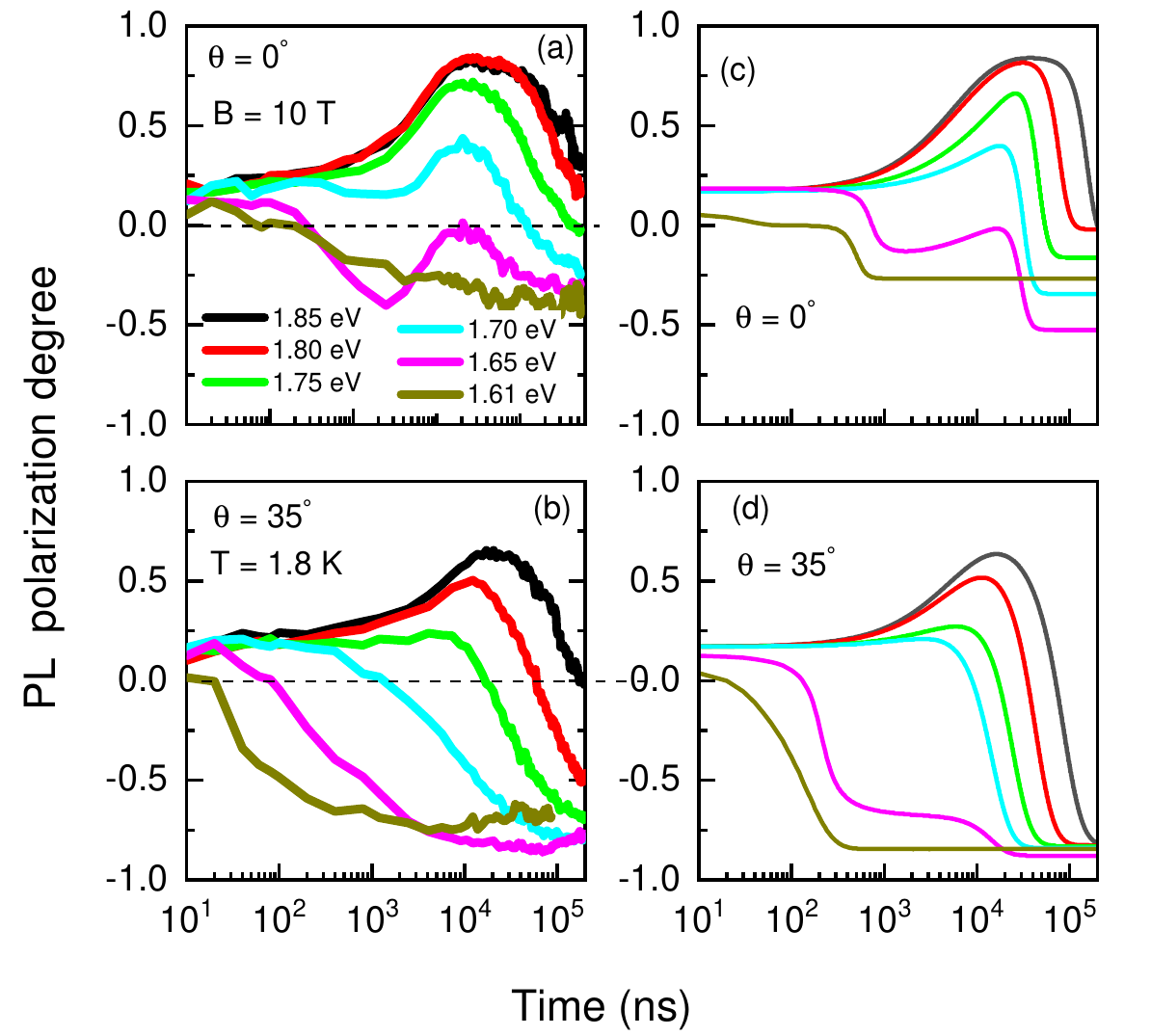}
\caption{\label{fig12} Collection of the dynamics of the magnetic-field-induced PL circular polarization degree measured along the PL band trace (a) in Faraday ($\theta=$ 0$^\circ$) and (b) tilted ($\theta=$ 35$^\circ$) geometries. $B=10$~T at $T=1.8$~K. Dynamics of $P_{c}(t)$, calculated using suggested theoretical model (c) in Faraday ($\theta=$ 0$^\circ$) and (d) tilted ($\theta=$ 35$^\circ$) geometries with parameters given in Sec.~\Ref{sec:fitting}.}
\end{figure}

The nonlinearity of the circular polarization degree dynamics is most clearly evident at high magnetic fields. To reveal thr features of $P_c(t)$ in QDs with different sizes, we measured them along the trace of the PL band for a longitudinal (Fig.~\ref{fig12}(a)) and tilted (Fig.~\ref{fig12}(b)) field of 10~T.

\subsection{Heavy hole and exciton $g$ factors in QD ensemble}

Recently, it was demonstrated that the spectral non-uniformity of the magnetic-field-induced polarization in a QD ensemble can be described by a model that takes into account a change of the sign of the bright exciton $g$ factor with changing QD size~\cite{Kotova,Kotova111}. In this model, it was also assumed that a relatively fast spin relaxation provides equilibrium populations on the long-living indirect exciton states, split by a magnetic field.

A non-monotonic dependence $P_c(t)$ in Faraday geometry allows one to conclude, that equilibrium populations of exciton states are not achieved for most of the QDs in the ensemble in a longitudinal field. However, the dependence of the exciton $g$ factor on (In,Al)As QD size still has not been finally established.

\begin{figure}[]
\includegraphics* [width=7.0cm]{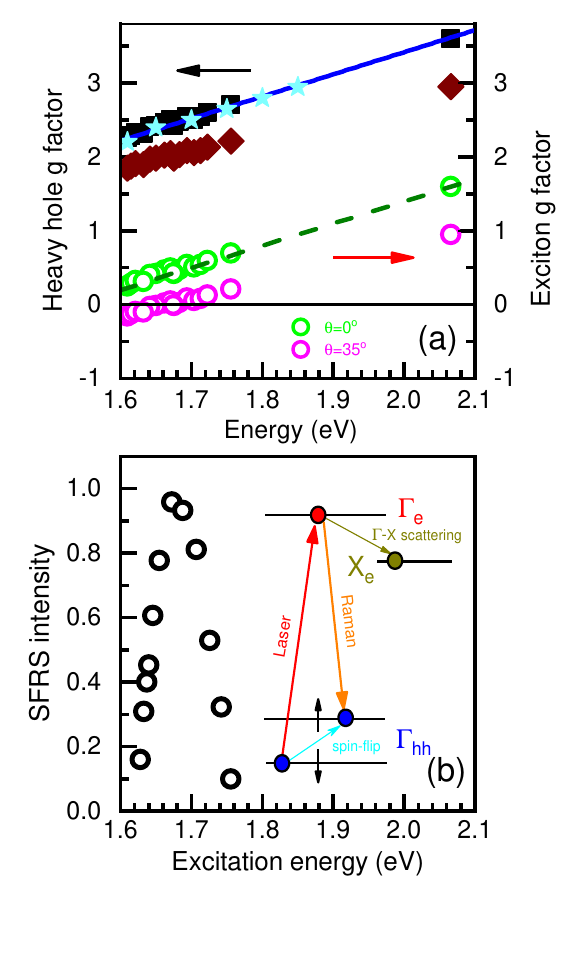}
\caption{\label{fig13}(a) Heavy hole longitudinal $\theta=$ 0$^\circ$ (black) and tilted $\theta=$ 35$^\circ$ (wine) $g$-factor $g_{hh}$, measured across the QD ensemble by spin-flip Raman scattering. Bright exciton longitudinal $\theta=$ 0$^\circ$ (green) and tilted $\theta=$ 35$^\circ$ (magenta) $g$ factor, calculated as $g_{b} =  g_{hh}- g_e$  with $g_e=+2$~\cite{Debus}. The point ($g_{hh} =+3.6$ was obtained in  Ref.~\cite{Shamirzaev104} for a  thin (In,Al)As/AlAs quantum  well  that  is  similar  to the  wetting  layer  in the  studied QD structures. Solid and dashed lines are linear approximations of the experimental data. The $g_{hh}$ values used in the calculation (Table~\ref{tab:parameters}) are shown by the cyan stars. (b) Intensity of heavy hole spin-flip Raman scattering signal as function of the excitation energy. Inset: Schematic diagram of the Raman spin-flip scattering process in the QDs, $\Gamma$-X scattering restricts the coherence time of the electron-hole pair. }
\end{figure}

In order to experimentally determine the exciton $g$ factor value as function of QD size, we applied the spin-flip Raman scattering technique, which can directly measure the Zeeman splinting of the electrons and heavy holes~\cite{Debus,Sirenko,Koudinov}. The fine structure of the involved states split by magnetic field is determined by the magnitude and sign of the heavy-hole ($g_{hh}$) and electron ($g_e$) $g$-factors that contribute the bright exciton $g$ factor, $g_{b} =  g_{hh}- g_e$. Due to the large band gap at the X point, the spin-orbit contribution to the electron $g$-factor is vanishingly small~\cite{ivchenko05a, vurg}. As a result, the electron $g$-factor  is isotropic and its value almost coincides with the free-electron Land\'{e} factor of +2.0~\cite{Debus,Ivanov97,Ivanov104}. Therefore, to determine $g_{b}$, the longitudinal ($\theta=$ 0$^\circ$, $g_{hh}$) and tilted $\theta=$ 35$^\circ$ heavy hole $g$ factor values within the QDs ensemble were measured. The obtained spectral dependencies of $g_{hh}$ and the calculated dependence of $g_{b}$  are shown in Fig.~\ref{fig13}(a). The additional point ($g_{hh} =+3.6$ in this figure was obtained in  Ref.~\cite{Shamirzaev104} for a thin (In,Al)As/AlAs quantum  well which is similar to the  wetting layer in the studied QD structures. The high-energy edge of the $g_{hh}$ dependence in the QD ensemble obtained in our measurements is restricted by the decrease of the heavy-hole Raman spin-flip transition intensity, see Fig.~\ref{fig13}(b). This  intensity decrease  is due to the shortening of the direct exciton coherence time in the intermediate $\Gamma$ state which results from the involvement of optical phonons in the scattering  process, caused by an increase in the efficiency of electron scattering from the $\Gamma$ to the X valley (shown schematically in the insert to Fig.~\ref{fig13}(b)) with increasinc splitting energy between these valleys in small size QDs.

Nevertheless, one can see from the obtained data that the values of the longitudinal $g_{hh}$  and, accordingly, $g_{b}$  factor increase monotonically with decreasing QD size. Moreover, the $g$ factors measured in Ref.~\cite{Shamirzaev104} for the limiting case of a one monolayer thick (In,Al)As sheet fit well to the linear approximation of the obtained data shown in Fig.~\ref{fig13}(a). Thus, the longitudinal $g_{hh}$ > +2 and $g_{b}$ is positive for the whole studied QD ensemble. The tilted $\theta=$ 35$^\circ$ bright exciton $g$ factor is negative for QDs emitting photons with energy $ < 1.7$~eV and positive for those emitting $>1.7$~eV.
\\

In conclusion of this section, we summarize the most important
experimental findings:

(1) The broad time-integrated PL spectrum of the (In,Al)As/AlAs QDs ensemble can be divided into three strongly overlapping bands with different characteristic recombination times: (i) the low-energy range of $1.55-1.65$~eV, dominating the spectrum for tens of nanoseconds after the excitation pulse, (ii) the high-energy range of  $1.7-1.9$~eV, dominating the spectrum after the first band has decayed for tens of microseconds after the excitation pulse, and (iii) the intermediate range of $1.6-1.7$~eV, which begins to dominate the spectrum a hundred microseconds after the end of the excitation pulse. The first band is due to exciton recombination in direct band gap QDs and excitons with strong $\Gamma$-X mixing~\cite{Rautert99, Rautert100, Nekrasov}, the other two bands are due to exciton recombination in different types of indirect band gap QDs.

(2) A magnetic field basically does not affect the unpolarized PL dynamics.

(3) In both longitudinal and tilted magnetic fields, within a few nanoseconds after the excitation pulse, the PL becomes circularly polarized across the entire energy range. In this time window the polarization degree $P^{0}_{c}$($B$) increases with increasing magnetic field strength and does not depend on the PL energy and the magnetic field orientation.

(4) The dynamics of the PL circular polarization degree, that reflect the dynamics of the charge carrier spin polarization in the QDs, are non-monotonic within times of up to milliseconds. The evolution of the PL circular polarization degree and sign strongly depends on the emission energy (determined by the QD size) and changes with magnetic field strength and orientation.

(5) The value of the longitudinal heavy hole $g$ factor  exceeds +2 in all QDs. Since the electron $g$ factor in the X valley is +2, the longitudinal $g$ factor of both bright $g_{b} =  g_{hh}- g_e$  and dark $g_{d} =  g_{hh} + g_e$ excitons in all indirect band gap QDs is positive.

\section{Model description}
\label{sec:discussion}

The obtained experimental data cannot be described within the framework of a two-level bright exciton scheme with quasi-equilibrium population of states which predicts a monotonically increasing dynamics for the magnetic-field inducted polarization with saturation in the long-time limit for magnetic fields of any orientation.

We will show that the change in the sign of the polarization along the PL spectrum and the non-monotonicity of the polarization degree dynamics can be described within the framework of a four-level, bright and dark exciton model and a redistribution of excitons between these states as result of spin relaxation, similar to indirect band gap QWs~\cite{Shamirzaev96}. In this section, we present a corresponding theoretical model and discuss the features of the recombination dynamics and spin dynamics in the QD ensemble.

\subsection{Theory}

We consider an exciton formed of a heavy hole in the $\Gamma$ valley of the (In,Al)As QDs and an electron in one of the X valleys. To model the time-dependent PL polarization degree in magnetic field, we apply the theory developed in Ref.~\cite{Shamirzaev96} with minor modifications. The circular polarization is determined by the optical transitions between exciton eigenstates. The exciton Hamiltonian in external magnetic field $\bf B$ has the form
  \begin{equation}
    \mathcal H=g_e\mu_{\tt{B}}\bf{B}\bf{s} + \frac{g_{hh}}{3} \mu_{\tt{B}}\tt{B_zj_z}.
  \end{equation}
Here $\bf{s}$ is the electron spin, $j_z=\pm 3/2$ is the heavy hole spin along the structure growth axis ($z$), $g_{e}$ and $g_{hh}$ are the electron and heavy hole $g$-factors, $\mu_B$ is the Bohr magneton, and $B_z=B\cos\theta$ with $\theta$ being the angle between magnetic field and $z$-axis. We neglect the electron-hole exchange interaction~\cite{Pikus} as compared with the Zeeman splittings and the transverse heavy hole $g$-factor~\cite{Marie1999,Debus}. We also neglect the possible intervalley electron scattering due to the hyperfine interaction~\cite{Avdeev2019}.

\begin{figure}[]
\includegraphics* [width=7.0cm]{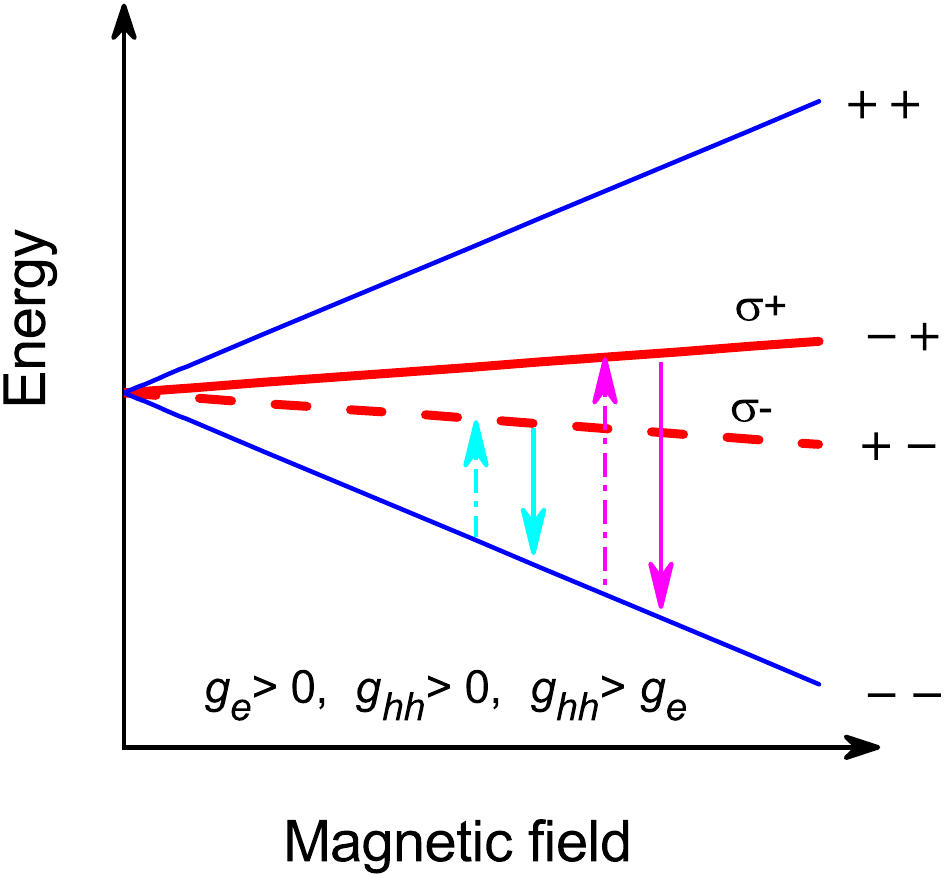}
\caption{\label{fig14} Schematic diagram of the exciton spin level structure in
a magnetic field, applied in the Faraday geometry, $\bf B \parallel z$. The case of positive $g_e$ and $g_{hh}$ with $g_{hh} > g_e$ is shown. The labels $++,--,-+$, and $+-$ correspond to the exciton states with spin projection $+2$, $-2$,
$+1$, and $-1$, respectively. The blue lines show the optically dark
(spin-forbidden) states. The red lines show the bright
(spin-allowed) states that result in $\sigma^+$ (solid line) or
$\sigma^-$ (dashed line) polarized emission. The arrows indicate
electron (hole) spin-flip processes increasing (dash-dotted) and
decreasing (solid) the carrier energy, respectively.}
\end{figure}

The eigenstates are characterized by the electron spin projection $\pm1/2$ on the direction of the magnetic field and the hole spin $j_z=\pm3/2$.
Coupling of the electron and heavy hole angular momenta provides an exciton fine structure that consists of four states. The two exciton states characterized by the angular momentum projections $\pm 1$ onto the $z$ axis are bright and the two states with the projections $\pm 2$ are dark (radiative recombination is forbidden by the spin selection rule for optical transitions)~\cite{Shamirzaev96,Shamirzaev94}. They are shown in Fig.~\ref{fig14}, and we denote the occupancies of these four states as $f_{\pm\pm}$ respectively. The occupancies satisfy the rate equations
  \begin{subequations}
    \label{eq:kin}
    \begin{equation}
      \frac{\d f_{++}}{\d t}=\frac{\alpha f_{-+}-f_{++}}{\tau_{se}}+\frac{\beta f_{+-}-f_{++}}{\tau_{sh}}-\frac{D^2}{\tau_r}f_{++},
    \end{equation}
    \begin{equation}
      \frac{\d f_{+-}}{\d t}=\frac{\alpha f_{--}-f_{+-}}{\tau_{se}}+\frac{f_{++}-\beta f_{+-}}{\tau_{sh}}-\frac{C^2}{\tau_r}f_{+-},
    \end{equation}
    \begin{equation}
      \frac{\d f_{-+}}{\d t}=\frac{f_{++}-\alpha f_{-+}}{\tau_{se}}+\frac{\beta f_{--}-f_{-+}}{\tau_{sh}}-\frac{C^2}{\tau_r}f_{-+},
    \end{equation}
    \begin{equation}
      \frac{\d f_{--}}{\d t}=\frac{f_{+-}-\alpha f_{--}}{\tau_{se}}+\frac{f_{-+}-\beta f_{--}}{\tau_{sh}}-\frac{D^2}{\tau_r}f_{--}.
    \end{equation}
  \end{subequations}
Here $C=\cos(\theta/2)$, $D=\sin(\theta/2)$, $\tau_e$ is the radiative recombination time, $\tau_{se,sh}$ are the electron and hole spin relaxation times,
  \begin{equation}
    \alpha=\exp\left(-\frac{g_e\mu_BB}{k_BT}\right),
    \qquad
    \beta=\exp\left(-\frac{g_{hh}\mu_BB_z}{k_BT}\right)
  \end{equation}
with $k_B$ being the Boltzmann constant, and $T$ the temperature. These equations were derived in Ref.~\onlinecite{Shamirzaev96} and the spin-flip related transitions are illustrated in Fig.~\ref{fig14}. We neglect the nonradiative recombination of excitons, which are strongly localized in QDs. As an extension of the model, we allow for initial spin polarizations of the electrons and holes, $P_e$ and $P_h$, respectively. This initial spin polarization may form during the energy relaxation of the charge carriers in the magnetic field after nonresonant optical excitation. It corresponds to the initial conditions for Eqs.~\eqref{eq:kin}
  \begin{equation}
    \label{eq:ics}
    f_{\sigma_e\sigma_h}=\frac{(1+\sigma_eP_e)}{2}\frac{(1+\sigma_hP_h)}{2}
  \end{equation}
with $\sigma_{e,h}=\pm$. The degree of PL circular polarization is then given by
  \begin{equation}
    \label{eq:pc}
P_c(t)=\xi\frac{C^{2}(f_{-+}-f_{+-})+D^{2}(f_{++}-f_{--})}{C^{2}(f_{-+}+f_{+-})+D^{2}(f_{++}+f_{--})}.
  \end{equation}
It can be calculated from the numerical solution of Eqs.~\eqref{eq:kin} with the initial conditions of Eq.~\eqref{eq:ics} and directly compared with the experimental results.

Note that the factor $D$ is responsible for the growth of the unpolarized PL intensity in tilted magnetic field, see Fig.~\ref{fig2}(b). In Eq.~\eqref{eq:pc}, we also introduced a phenomenological depolarization factor $0 < \xi < 1$~\cite{Shamirzaev96}, accounting for the deviation of the selection rules from the pure ones and also taking into account the loss of polarization of emitted light during its propagation through the structure.
In the Faraday geometry, the factors $C = 1$ and $D = 0$ and the dark exciton does not directly radiatively recombine. However, as was shown recently, the long living dark states can become weakly optically active due to minor deviations from the selection rules~\cite{Shamirzaev96}. Taking into account the dark-state activation factor $C_d$, we define the degree of exciton PL circular polarization in the Faraday geometry, as follows:
\begin{equation}
    \label{eq:pcf}
    P_c(t)=\xi\frac{(f_{-+}-f_{+-})+C_d(f_{++}-f_{--})}{(f_{-+}+f_{+-})+C_d(f_{++}+f_{--})}.
  \end{equation}

\begin{table*}[]
    \caption{Parameters of the studied (In,Al)As/AlAs QDs evaluated from best fits to the experimental data in magnetic field of 10~T.}
  \begin{ruledtabular}
        \begin{tabular}{lccccccc}
           Parameter~$\setminus$~PL energy &  1.61~eV    &     1.65~eV        &~1.70~eV      &  ~1.75~eV  &  1.80~eV    &  1.85~eV      &Comment      \\\hline
            $g_{e}$                        & $+2.0$      & $+2.0$             & $+2.0$       & $+2.0$     & $+2.0$      & $+2.0$        &  \cite{Debus,Ivanov97,Ivanov104} \\
            $g_{hh}$                       & $+2.2$      & $+2.4$             & $+2.5$       & $+2.65$    & $+2.8$      & $+2.95$       & Fig.~~\ref{fig13}(a)\\
            $\tau_{r}$                     & 10~ns       & 4.5$~\mu$s/80~ns   & 5.5$~\mu$s   & 12$~\mu$s  & 15$~\mu$s   & 30$~\mu$s     & best fit\\
            $\tau_{sh}$                    & 70~ns       & 4.5$~\mu$s/110~ns  & $5$~$\mu$s   & $7$~$\mu$s & $22$~$\mu$s & $50$~$\mu$s   & best fit\\
            $\tau_{se}$                    & 70~ns       & 3$~\mu$s/100~ns    & $3$~$\mu$s   & $3$~$\mu$s & $3$~$\mu$s  & $3$~$\mu$s    & best fit\\
            $\xi$                          & $0.85$      & $0.85$             & $0.85$       & $0.85$     & $0.85$      &  $0.85$       & best fit\\
            $P^{0}_{c}$                    &   0.1       &       0.2          &      0.2     &      0.2   &   0.2       &  0.2          & Figs.~\ref{fig10}~-~\ref{fig11}  \\
            $C_d$                          &        0    &     0.0002         &    0.0002    &   0.0006   &   0.001     &    0.001      & best fit\\
           Time-integrated  $P_c$          & negative    &   negative         &  positive    &  positive  &    positive &  positive     &     \\
            \end{tabular}
  \end{ruledtabular}
    \label{tab:parameters}
\end{table*}

\subsection{Fitting the experimental data}
\label{sec:fitting}

Admittedly, the model has a fair number of parameters. The photoluminescence intensity and  polarization degree are governed by the following factors: the values and signs of the electron and hole $g$ factors, the magnetic field strength and orientation, the radiative recombination times, the spin relaxation rates describing the spin flip rates for downward transitions, i.e., from the upper to the lower Zeeman sublevel, and the temperature, which determines the ratio of upward and downward transitions. Some of these parameters can be directly measured in experiment, others can be evaluated from fits to the various experimental dependencies, or at least their possible value ranges can be  directly measured in experiment. As a result, we obtain quite stringent estimates of the parameter values.

Let us start from well defined parameters. Due to the large band gap at the X point, the spin-orbit contribution to the electron $g$-factor is vanishingly small~\cite{ivchenko05a}. As a result, the electron $g$-factor, $g_e$, is isotropic and its value almost coincides with the free-electron Land\'{e} factor of
+2.0~\cite{Debus,Ivanov97,Ivanov104} for QDs of any size. The heavy hole $g$-factor is strongly anisotropic. However, it was shown that the transverse heavy hole $g$-factor in (In,Al)As/AlAs QDs is very close to zero $g^{\perp}_{hh} =0.03\pm0.05$~\cite{Debus}. Therefore, we can take the angular-dependence of the heavy hole $g$ factor as $g_{hh}(\theta) = g_{hh} \cos \theta $, and the dependence of the longitudinal $g_{hh}$ factor on QD size is given by fitting the SFRS data presented in Fig.~\ref{fig13}(a).  As for the exciton recombination time, in contrast to the case of QWs, where the exciton recombination times can be found directly from exponential PL decay measurements~\cite{Shamirzaev96,Shamirzaev104,Shamirzaev106}, the situation in QD ensembles is more complicated. The exciton recombination dynamics show a nonexponential decay as a result of the superposition of multiple monoexponential PL decays with different lifetimes due to the dispertion of QD/matrix interface  sharpness for different QDs, which emit photons with the same energy~\cite{Shamirzaev84}. Therefore, the bright exciton lifetime is taken as a fit parameter. Note, that direct measurement of the PL dynamics in direct band gap QDs revealed an exciton lifetime of about 1~ns~\cite{Rautert100}, therefore, the spin dynamics with times much longer than 1~ns in low-energy spectral range, see Fig.~\ref{fig12}(b), arises from indirect band gap QDs with strong $\Gamma$-X mixing.

In total, we have used five variable fit parameters to describe the experimental findings:  the bright exciton lifetime $\tau_{r}$, the electron $\tau_{se}$ and heavy hole $\tau_{sh}$  spin relaxation times, the depolarization factor ${\xi}$ and the dark-state activation factor $C_d$, the two last parameters accounting for a small deviation of the selection rules from the pure ones~\cite{Shamirzaev96}. However, the possibilities for variation of ${\xi}$ and $C_d$ are quite limited. As a reference value for ${\xi}$ we take the saturation level of $P_c$ in magnetic field in Fig.~\ref{fig8}, and $C_d \ll 1$~\cite{Shamirzaev96}.

Modeling the experimental data by kinetic master equations, we also have to take into account some initial polarization $P^{0}_{c}$ that does not depend on the PL energy and magnetic field orientation. Recently this was observed in the appearance of circularly polarized emission immediately after the end of the excitation pulse in another indirect band gap system,  Ga(Sb,P)/GaP~\cite{ShamirzaevJL}. This feature is determined by the fast spin polarization of electrons and holes in the matrix. These spin-polarized charge carriers are captured from the matrix into the QDs, so that some initial spin polarization of QD excitons occurs. We take this polarization into account through the initial conditions for solving the kinetic master equations. Note that these initial conditions are unambiguously determined from the experimental data (Figs.~\ref{fig10}~-~\ref{fig11}).
Prominent characteristic features of the $P_{c}$ dynamics along the PL band trace appear for longitudinal and tilted magnetic fields of 10~T, as presented in Figs.~\ref{fig12}(a) and~\ref{fig12}(b), respectively.  Therefore, we focus in model description of these experimental data.

The comparison between measurements and calculations done using the suggested theoretical model is shown in Fig.~\ref{fig12}. The parameters obtained from the best fit of the  experimental data are collected in Table~\ref{tab:parameters}. The same set of parameters was used for both the longitudinal  ($\theta=0^{\circ}$) and tilted ($\theta=35^{\circ}$) magnetic fields. One can see that the model allows us to get good correspondence between the measured and calculated dependencies. Some difference is related to the inhomogeneity of the quantum dot parameters (mainly, due to exciton lifetimes dispersion) that contributes to the emission at a certain energy~\cite{Shamirzaev84}. In the high energy spectral range (above 1.70~eV) the dynamics are described by a single set of parameters that change monotonically with QD size change. In the considered QDs, the carriers are characterized by a "slow" spin relaxation occuring during times in the microsecond region. A quite different set of parameters is found for the low energy spectral range (1.61~eV), which corresponds to larger size QDs with $\Gamma-$X mixing of the electron states which results in "fast" exciton recombination and carrier spin dynamics. Both types of QDs with "slow" and "fast" spin dynamics contribute to the intermediate range (1.65~eV).

\section{Discussion}
\label{sec:disc}

Let us first discuss the exciton lifetime as well as electron and heavy hole spin relaxation time dependences on the QD size. The sub-ensemble of large-size (In,Al)As/AlAs QDs is characterised by "fast" exciton recombination and short spin relaxation times in the range of several tens of nanoseconds. These values are in good agrement with the results of our previous studies~\cite{Shamirzaev84,Nekrasov}. Indeed, the small energy difference between the $\Gamma$ and X electron states in these QDs provides sizable $\Gamma-$X mixing which leads to: (1) a short exciton recombination time~\cite{Shamirzaev84} that approaches the direct exciton lifetime of 1 ns, and (2) an increase of the exciton anisotropic exchange interaction that accelerates spin relaxation~\cite{Nekrasov}. The "short" spin relaxation times in QDs with strong $\Gamma-$X mixing result in a quasi-equilibrium population of the Zeeman sublevels resulting in the  well known scenario, where the positive $g$ factor of the bright exciton leads to a negative magnetic-field-induced polarization of the exciton emission~\cite{Ivchenko60} as observed in this study.

In small-size indirect band gap QDs with small $\Gamma-$X mixing the exciton lifetime is determined by the QD/matrix heterointerface sharpness~\cite{Shamirzaev84} and one can increase the exciton lifetime up to hundreds of microseconds, in good agreement with values from model calculations. In these small size QDs, the spin relaxation times are of comparable magnitude and for a sub-ensemble can be even longer than the exciton lifetimes. Therefore, a quasi-equilibrium population of the Zeeman sublevels is not achieved and the polarization of the exciton emission is mainly given by the ratios between the different relaxation times in the system~\cite{Shamirzaev96}. Since the ration between the longitudinal electron and heavy hole $g$ factors $g_{hh} > g_{e}$ >0 is maintained withing the whole ensemble of QDs, a positive polarization of the exciton emission unambiguously indicates that the heavy hole spin relaxation time is longer than that of the electrons~\cite{Shamirzaev96}. That agrees well with the fitting result and corresponds to the theoretical study by Bulaev and Loss~\cite{Bulaev2005}, which predicts spin relaxation times for heavy-holes comparable to or even longer than those for electrons in two-dimensional QDs. We believe that the heavy hole spin relaxation time is mainly determined by the heavy-light holes coupling. Therefore, an increase of $\tau_{sh}$  with decreasing QD size is as result of the increase of the heavy-light holes energy gap with increasing quantum confinement.

It is interesting that electron spin relaxation time does not change with QDs size, while for the heavy hole it increases with decreasing QD size. Previously, it was shown that at low temperatures the electron spin relaxation rate in direct band gap III-V structures depends strongly on size, temperature and magnetic field \cite{Woods2002}. For temperatures in the range of a few degrees Kelvin, the electron spin relaxation rates are given by single-phonon emission processes accompanied by spin flips arising from spin-orbit coupling~\cite{Tsitsishvili2003}. The constant electron spin relaxation rates in indirect band gap QDs of any size can arise from negligible spin-orbit coupling in the X value of the conduction band. However, further studies need to clarify this feature.

Let us discuss the difference in spin dynamics between thin QWs~\cite{Shamirzaev104} and QDs that formed in (In,Al)As/AlAs heterostructures. In both cases the exciton spin dynamics can be described by a four-level model that takes into account the exciton redistribution on the bright and dark states. The model describes well the non-monotonic dynamics of the magnetic field induced PL polarization degree in longitudinal and tilted fields. However, there are some differences among which the most striking is the dependence of the unpolarised PL intensity and PL dynamics on longitudinal magnetic field. Indeed, the unpolarised PL intensity decreases by a factor of five and the PL decay time increases from 0.21~ms to 0.45~ms with increasing longitudinal magnetic field from 0 up to 10~T in the case of  QWs~\cite{Shamirzaev104}. In contrast, in the case of QDs the PL intensity and PL decay basically do not change with magnetic field. It was previously demonstrated that in thin QWs the difference in PL intensity and decay at zero and high magnetic field increases with increasing non radiative recombination time of the dark exciton states~\cite{Shamirzaev104,Shamirzaev96}. However, indirect evidence suggests that the dark exciton nonradiative time in (In,Al)As/AlAs QDs has to be longer than that in thin QWs due to strong localization of excitons in the QDs. In QWs, the PL intensity monotonically decreases by about one order of magnitude when the temperature increases from 2 K up to 15 K~\cite{Shamirzaev104}. This decrease is related to nonradiative centers at the heterointerfaces, which capture  excitons  more  efficiently  at  elevated  temperatures, because the exciton diffusion length simultaneously increases. On the other hand, the PL intensity in QDs is almost constant up to room temperatures~\cite{ShamirzaevJETP}. Moreover, in spite of neglecting the non-radiative recombination processes we obtain a good description of the experimental polarization dynamics by the model calculations.

The exciton emission in thin indirect band gap QWs and indirect band gap QDs is provided by quite different mechanisms. In the case of the QWs it is given by phonon assisted recombination, while for the QDs the emission is provided by no-phonon recombination via  $\Gamma-$X mixing at the QDs/matrix heretojunction. We believe that scattering at the QD heterointerface can provide some additional mixing of bright and dark exciton states that leads to maintain the unpolarized PL dynamics when changing the magnetic field. In order to clarify this feature we plan to extend our studies of long-lived localized excitons to other indirect band gap QDs.

\section{Conclusions}
\label{sec:conclusions}

The recombination and spin dynamics of excitons that are indirect in momentum space and involve electrons from the X valley were investigated in an ensemble of (In,Al)As/AlAs quantum dots with type I band alignment. We  demonstrate that in spite of the long exciton lifetime, the magnetic field-induced circular polarization of the exciton PL in these QDs is not completely controlled by the thermodynamic parameters, i.e., the ratio of the exciton Zeeman  splitting  and  the  thermal  energy does not provide a quasi-equilibrium population of the exciton states. The spin dynamics in these QDs are determined by kinetic parameters, i.e., the relations between the various relaxation times in the system, which depend on the QD size. A characteristic feature of the (In,Al)As/AlAs QDs is the fast spin polarization in strong magnetic fields for electrons and holes during energy relaxation in the matrix. The capture of these spin polarized carriers into the QDs provides some initial spin polarization of excitons on times much faster than the electron and hole spin-relaxation times in the QDs.

{\bf Acknowledgements.} We thank \href{https://orcid.org/0000-0003-4462-0749}{M. M. Glazov} for fruitful discussions. The experimental activities   conducted   by   T.S.S.,   including   sample   growth, investigation   of   energy   level   spectra   and   magneto-optical   properties, as   well   as   exciton   recombination   and   spin   dynamics,   were   supported   by a   grant   of the  Russian   Science   Foundation   (project   No.   22-12-00022-P).  The theoretical work by V.N.M, was supported by the Foundation for the Advancement of Theoretical Physics and Mathematics “BASIS”. D.R.Y, D.K., and M.B. acknowledge the financial support by the Deutsche Forschungsgemeinschaft through the Collaborative Research Center TRR142 (Project A11).

\end{document}